# Surfactants Screen Slide Electrification

Xiaomei Li* [a][b], Zhongyuan Ni[a], Xiaoteng Zhou[a], Lisa S. Bauer[c], Diego Diaz[a], Gabriele Schäfer[a], Hans-Jürgen Butt* [a]

[a] Max Planck Institute for Polymer Research, Ackermannweg 10, 55128 Mainz, Germany

[b] ETH Zürich, Department of Chemistry and Applied Biosciences, 8093 Zürich, Switzerland

[c] TU Darmstadt, Institute for Nano- and Microfluidics, Peter-Grünberg-Straße 10, D-64287 Darmstadt, Germany

E-mail:  xiaomei.li@chem.ethz.ch and butt@mpip-mainz.mpg.de

**Abstract**: *Water drops spontaneously accumulate charges when they move on hydrophobic dielectric surfaces by slide electrification. On the one hand, slide electrification generates electricity with possible applications on tiny devices. On the other hand, the potential of up to 1 KV generated by slide electrification alters wetting and drop motion. Therefore, it is important to know the factors that affect slide electrification. To find out how surfactants affect slide electrification, we measured drop charges of aqueous drops containing cationic CTAB, anionic SDS and neutral $C_8E_3$ sliding on different hydrophobic surfaces. The result is: addition of surfactant significantly reduces the spontaneous charging of moving water drops. Based on zeta potential measurements, confocal microscopy of deposited surface-active dyes and drop impact studies, we propose that several factors contribute to this suppression of charge separation: (1) Surfactants tend to lower the contact angles, which reduces charge separation. (2) Surfactant adsorption at the solid-liquid interface can reduce the density of primary ions, particularly for anionic surfactants. (3) Anionic and neutral surfactants are mostly transferred to the liquid-air interface at the rear of the sliding drop, retaining primary ions within the drop. (4) Deposited cationic surfactant directly reduces the charge of the drop.*

When aqueous drops de-wet hydrophobic solid surfaces, charges spontaneously accumulate inside the drops and leave an opposite charged surface behind. This process is termed "slide electrification" or "contact electrification" [1,2]. Such charge separation can be used to harvest electricity for small-scale devices applicable in remote and off-grid areas, to manipulate drop motion, or to modify surface properties. For instance, Xu *et al.* designed a drop-based electricity generator with high instantaneous power density, which can light up commercial LEDs [3]; Sun *et al.* printed charges on superhydrophobic surfaces with bouncing drop for programmed drop transport [4]. On the other hand, slide electrification contributes substantially to drop friction [5] or contact angle hysteresis [6]. Drop potentials can reach more than 1 KV [7-9]. Such high potentials may lead to corrosion. For this reason, it is important to clarify under which conditions slide electrification occurs and how to control slide electrification and its influence.

A general strategy to control contact electrification is surface modification. For example, by coupling the electron-donating and the electron-accepting component on solid surfaces, Zhang *et al.* designed non-charging surfaces from non-conductive polymers [10] and Jin *et al.* eliminated solid-solid contact electrification [11]. By functionalizing hydrophobic surfaces with protons accepted amine-components, Wong *et al.* also succeed to tune the charge of sliding water drops on hydrophobic surfaces [12]. In addition to the modification of solid surfaces, increasing evidences indicate that the components of liquid (salt and pH) also have significant impacts on the charge separation during slide electrification. For example, Helseth showed the existence of an optimal salt concentration which allows maximum charge transfer when a water front moves across a junction between a hydrophobic dielectric and a metal electrode [13]. Sosa *et al.* reported a point of zero charge at pH=3 for sliding water drops on PTFE surfaces [14]. Thus, it is crucial to know the role of liquid components to control charge separation either for energy harvest or corrosion prevention. Since water is naturally contaminated, many contaminants strongly enrich at the water's surface and act as surfactants, we need to know the influence of surfactants to estimate how sensitively charge generation depends on clean water or possible contamination.

**Table 1** *Critical micelle concentration (CMC), surface tension ($\gamma$) at CMC, diffusion coefficient in water (D) at 25 ℃ for the three surfactants.*

| Surfactant | Type | CMC (mM) | $\gamma$ (mNm$^{-1}$) | $D$ (m$^2$s$^{-1}$) |
|---|---|---|---|---|
| CTAB [15] | Cationic | 1.0 | 35.2 | $5.4 \times 10^{-10}$ |
| SDS [16] | Anionic | 8.1 | 33.5 | $2.5 \times 10^{-10}$ |
| C$_8$E$_3$ [17] | Nonionic | 7.5 | 27.3 | $4.6 \times 10^{-10}$ |

To evaluate the impact of surfactant on slide electrification, we measured the charge of water drops with cationic, anionic, and nonionic surfactants at comparable critical micelle concentration (CMC) sliding on hydrophobic surfaces. As surfactants, we chose cetyltrimethylammonium bromide (CTAB, >98.0%, Tokyo Chemical Industry Co., Japan), sodium dodecyl sulfate (SDS, >99%, MP Biomedicals Germany GmbH), and *n*-octyltrioxyethylene (C$_8$E$_3$, BACHEM, Switzerland) (Table 1, Supporting information SI 1). The setup for the measurement of drop charge was home-built by coupling a tilted plate and a current amplifier (Figure 1a) [2]. During charge measurement, an ionizing air blower was running in front of the samples. The ionic blowing was strong enough to neutralize the surface between two drops in tens of seconds but so weak that it did not significantly reduce drop charging while the drop was running down the tilted plate in tens of milliseconds. The syringe size was fixed for all solutions. With increasing surfactant concentration, the drop volume decreased from ~45 µL to ~25 µL for ionic surfactants and from ~45 µL to ~16 µL for $C_8E_3$. We detected the drop current after drops sliding for 4 cm on hydrophobic surfaces by an electrode connected to a current amplifier (SI 2). By integrating the current signal over time during which the peak current appears (typically ~2 ms for a water drop), the cumulative drop charge is obtained. Four different hydrophobic coatings on quartz substrates (76.2×25.4×1.0 mm$^3$) were measured: (1) monolayers 1H, 1H, 2H, 2H-perfluorooctadecyltrichlorosilane (PFOTS) prepared by chemical vapor deposition; (2) polydimethyl-siloxane brushes (PDMS) prepared by immersion for 24 h into the melt ($M_w$ = 6 kg/mol) at room temperature and excessive rinsing with toluene; (3) 30±5 nm polystyrene films (PS) prepared by dip-coating from a toluene solution; (4) 60±5 nm Teflon AF1600 thin film (Teflon) also prepared by dip-coating. Teflon films were also formed on glass slides coated with 5 nm chromium plus 35 nm gold layer. The preparation details are referred to (SI 1) [5]. We used fresh surfaces for each measurement to avoid the influence of deposited surfactant on the surfaces from previous measurements.

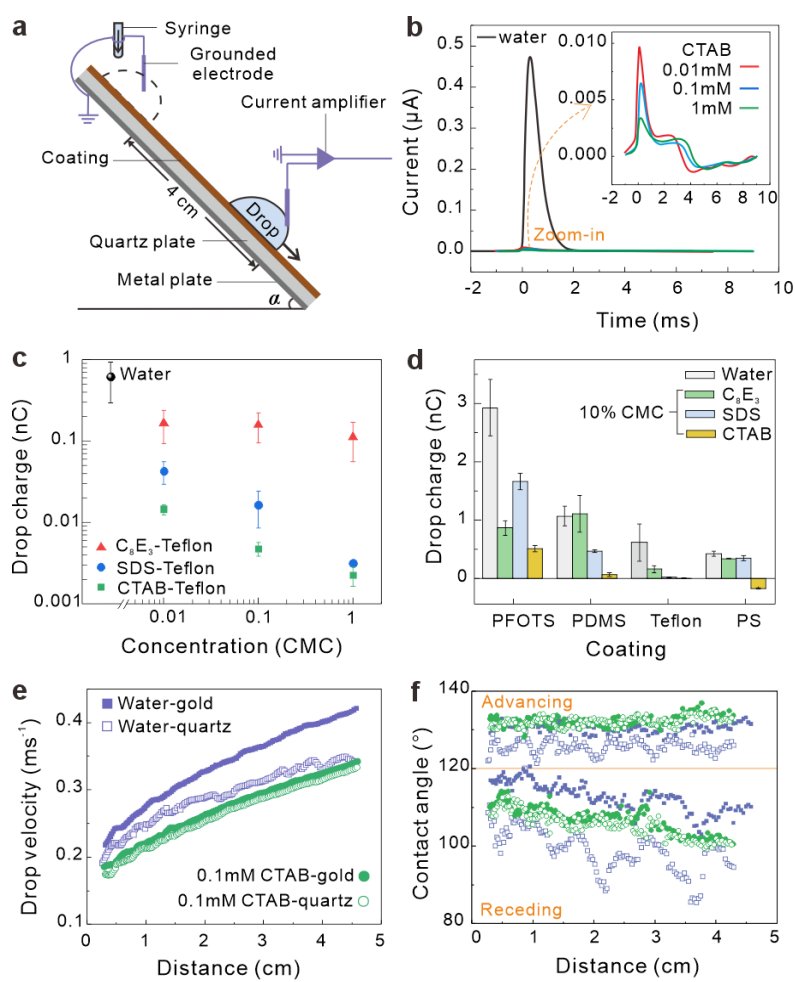

**Figure 1.** *Screening effect of surfactant on charging of sliding drops. (a) The tilted plate setup for charge measurement and high-speed video recording from the side. To avoid the influence of noise from the*

*high-speed camera on charge measurement, both measurements were carried out separately. (b) Drop currents of water drops containing 0, 0.01, 0.1 and 1 mM CTAB sliding on Teflon-quartz surfaces. (c) Drop charge versus concentration of water drops with different surfactants sliding down Teflon-quartz surfaces. The concentrations were normalized by division with the respective CMC. (d) Drop charge of water drop with 10% CMC CTAB, SDS, and $C_8E_3$ on different surfaces. The tilted angle of the surface was 50° for PFOTS-quartz and PDMS-quartz samples, 40° for PS-quartz, and 25° for Teflon-quartz. (e) Drop velocity and (f) dynamic contact angles versus slide distance of water drops with 0 mM and 0.1 mM CTAB on Teflon-quartz and Teflon-gold surfaces.*

When adding surfactant and increasing its concentration, the drop current decreased substantially (Figure S1). One representative example measured on a Teflon-quartz surface is shown in Figure 1b. The peak current is reduced by more than a factor of 10 when 0.01 mM CTAB was added (1% CMC). Integrating the peak current over the corresponding time gives the drop charge. Drop charges decreased with increasing normalized surfactant concentration for all three of surfactants (Figure 1c). We normalized the concentrations in mol/L by the respective CMC. The CMC is a measure of the strength of the hydrophobic interaction between the tails of the surfactants. The same hydrophobic interaction leads to adsorption at the water-air and the hydrophobic solid-water interfaces. It is therefore a common scaling factor. The strongest reduction in drop charge was observed with CTAB followed by SDS. The effect of nonionic surfactant was less strong. The reduction of drop charge by adding surfactant occurs for all tested hydrophobic surfaces (Figure 1d). The charge signal even reversed from positive to negative for water drops with 10% CMC CTAB sliding on a PS-quartz surface (Figure 1d).

Addition of surfactant also decreased the velocity of drops (Figure 1e, solid points). This reduction in velocity is mainly due to the decrease of receding contact angles (Figure 1f), which results in an increase in the contact angle hysteresis, consistent with previous report [15]. The effect was explained with a Marangoni stress caused by the gradient in surface excess of surfactants at the rear of drops. SDS and $C_8E_3$ also have similar screen effect on drop motion (Figure S2).

The drop charge is determined by the deposited surface charge density and the dewetted surface area. Although lower contact angles lead to larger solid-liquid contact areas, the slight increase is however compensated by a decreasing drop size with surfactant. The surface charge is governed by three different effects [18]: surface chemistry, contact angle, and fluid flow. Surface chemistry determines the surface charge density ($\sigma_{eq}$) in the electric double layer, which can potentially be separated during dewetting. Contact angle ($\theta$) mainly affects countercharge distribution near contact line, local surface charge near contact line increases for $\theta > 45°$ while decreases for $\theta < 45°$. The impact of fluid flow is quantified by the Peclet number $Pe = \frac{U\lambda}{D}$ ($U$: contact-line velocity, $\lambda$: Debye length). Coupling the three effects, the deposited surface charge density ($\sigma_s$) can be estimated by $\sigma_s \propto \frac{\sigma_{eq}\, \theta}{Pe}$. Therefore, to understand why and how surfactants hinder slide electrification, we consider several mechanisms:

1) Surfactants reduce both liquid-air ($\gamma_L$) and liquid-solid interfacial energy ($\gamma_{SL}$) so that contact angles decrease according to Young's equation, $\cos\theta = (\gamma_S - \gamma_{SL})/\gamma_L$ ($\gamma_S$: solid surface energy) [19]. The reduction of static contact angles by additive surfactants with different concentrations is listed in Table S1. Lower contact angles reduce charging, in line with the charge results. However, since the nonionic surfactant $C_8E_3$ showed the strongest decrease in contact angles but had only a weakly reduced drop charging, we exclude this effect as the main mechanism.

2) Surfactants affect the fluid flow. Additional surfactants reduce liquid diffusivity and contact-line velocity. The two effects will partially offset each other and have less influence on charging. Moreover, for ionic surfactants, the Debye length decreases from typically 200-300 nm for distilled water to 10 nm at 1 mM and 4 nm at 8 mM ionic surfactant, the decreasing Debye length is expected to decrease $Pe$ and increase drop charge, which counteract to the observed decrease in drop charge [7, 18]. Thus, the effect of fluid flow should not be the dominant mechanism neither.

3) Surfactant absorption changes surface chemistry. Surfactants adsorb to the hydrophobic solid-liquid interface due to hydrophobic bonding between surfactant tails and hydrophobic surfaces. Since hydrophobic surfaces normally tend to be charged negatively in contact with water [20-24], cationic CTAB can further absorb to the solid-liquid interface by Coulombic attraction. Despite Coulombic repulsion, anionic SDS absorbing on the negatively charged hydrophobic surfaces is still possible when the hydrophobic-bonding force overcomes the Coulombic repulsion [25]. This adsorption may change the surface chemistry and therefore surface charge density ($\sigma_{eq}$) in the electric double layer.

4) For ionic surfactant, there may be a direct charge deposition by remaining on the solid surface behind the drop. Charged surfactant could remain adsorbed to the solid surfaces or could be transferred at the rear contact line from the bulk driven by the increasing potential in the moving drop.

To gain more insight into adsorption of surfactant to the solid-liquid interface, we measured zeta potentials at the interface between aqueous solutions with 100% CMC surfactants concentration and our hydrophobic surfaces (SI 4). The zeta potentials were derived from streaming potentials and streaming currents using an electrokinetic analyzer (SurPASS 3, Anton Parr, Austria) at pH around 5.5-6.0. We used 0.5 mM KCl aqueous solution as a reference, the zeta potential was around -30 mV for different hydrophobic surfaces (Figure 2a, grey). This value agrees with the zeta potential reported in the literature [26, 27]. The negative zeta potential of hydrophobic surfaces in salt solutions is usually interpreted by an enrichment of hydroxyl groups [28-30]. Here, we call the ions which dominate the surface potential in the absence of surfactant primary ions. With 100% CMC CTAB, the zeta potential reversed sign and changed to +60 mV for all the tested hydrophobic surfaces (Figure 2a, yellow). In 100% CMC SDS, the zeta potential became even more negative than the reference KCl solution, about -60 mV (Figure 2a, blue). 100% CMC $C_8E_3$ slightly reduced the zeta potential to about -22 mV (Figure 2a, green).

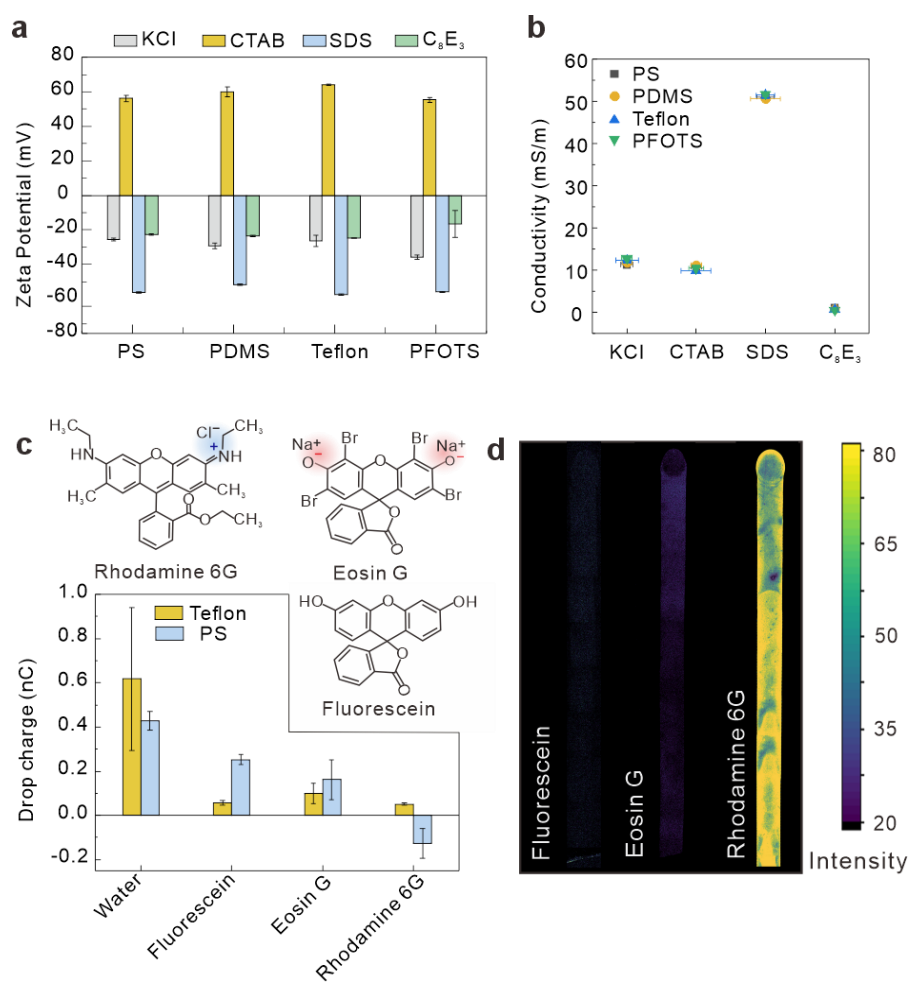

**Figure 2.** *Proof of concept of the mechanism. (a) Zeta potential and (b) conductivity of aqueous solutions with 0.5 mM KCl or 100% CMC surfactant at different hydrophobic surfaces on quartz substrates. (c) Drop charge of water with 0.1 mM fluorescein, eosin G, and rhodamine 6G sliding on the Teflon-quartz surfaces and PS-quartz surfaces. (d) Confocal images of drop paths with ~6 cm length and 6 mm width after water drops with fluorescein, eosin G, and rhodamine 6G at a concentration of 0.1mM sliding on the fresh PDMS-quartz surfaces.*

When adding surfactants, the conductivity of the liquid changes as well (Figure 2b). To avoid the influence of liquid conductivity on zeta potential measurements, we unified liquid conductivity by using same surfactant concentration and adding additional 0.5 mM KCl in $C_8E_3$ solution. In this control measurement, the trends of zeta potential (Figure S3) are identical to those in the Figure 2a. Similar changes in zeta potential have been reported on hydrophobic surfaces [28-30]. The change of zeta potential by surfactants demonstrates the absorption of surfactant at the solid-liquid interface.

An important question is whether absorbed surfactant remains on the solid surface after dewetting, which would directly imply a charge or surfactant deposition. To answer this question, we used fluorescent dyes to mimic surfactants. By measuring the fluorescence intensity of the sliding drop path on solid surfaces with confocal microscopy after dewetting, we determined the intensity of deposited dye on the solid surface (SI 5). Three dyes with diffusion coefficient of $\sim 4.0 \times 10^{-10}$ $m^2s^{-1}$ were used: cationic Rhodamine 6G, anionic Eosin G, and nonionic Fluorescein [31-33]. The charge of water drops with 0.1 mM (around 10% CMC) dye was identical to the case of water drop with 10% surfactants (Figure 2c). We presume that the deposition of dyes on the hydrophobic surfaces is comparable to the deposition of surfactants.

The fluorescence images in Figure 2d show the drop paths after a water drop with 0.1 mM dye sliding down a pristine PDMS-coated quartz surface. The fluorescence intensity of drop path with cationic Rhodamine 6G was much larger than the other two, indicating more Rhodamine 6G was deposited on the surface after dewetting (Figure 2d and Figure S4). Based on the reduction of drop charge by Rhodamine 6G compared with pure water (Figure 2c), we estimated a mean spacing between deposited Rhodamine 6G molecules of ≈200 nm (SI 8). Cationic dye deposited on the surface could be due to the absorbed dye at the liquid-solid interface un-desorbed when contact-line recedes or it is transferred at the rear of contact line from the bulk driven by the increasing potential in a moving drop. Almost no anionic and nonionic surfactant dye residues appear after dewetting. This indicates that the pre-absorbed surfactant at the solid-liquid interface desorbed again by receding contact line due to the capillary force [34].

Based on the experimental observations, we interpret the reduction of charge separation by surfactants with a two-step model. First, surfactants are adsorbed at the liquid-solid interface (Figure 3, pathway 1). Surfactant adsorption is verified by zeta potential measurements. Adsorption is driven by hydrophobic interactions and either enhanced (for CTAB) or weakened (for SDS) by electrostatic interactions. It is well established that cationic, neutral, and anionic charged surfactants adsorb to hydrophobic surfaces [35-39]. Even at concentrations below the CMC, a large part of the solid surface is shielded by the surfactant. As a result, fewer primary ions can bind to the solid surface.

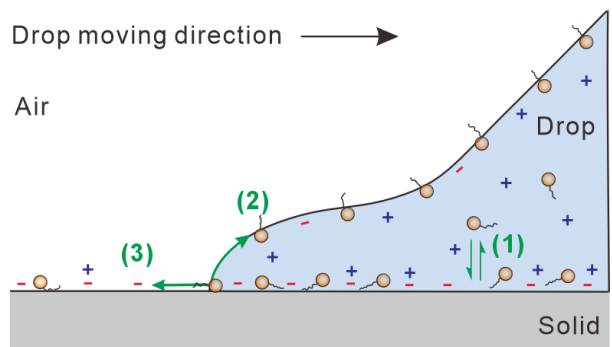

**Figure 3.** *Mechanisms for the screen effect of surfactant on slide electrification. (1) Surfactants and ions reversibly absorb to the solid-liquid interface, which changes the surface chemistry; (2) Part of the physiosorbed surfactants desorb at receding contact line due to capillary force during dewetting; (3) The deposition of primary ions and surfactant to the solid surface after dewetting leads to the net surface charge. The red minus signs indicate the primary ions.*

The second step takes place at the rear of the contact line. Most of the surfactant is transferred to the water-air interface (Figure 3, pathway 2) or diffuses back into the bulk. Only a tiny fraction of the adsorbed surfactant is deposited behind the drop during dewetting. This is similar to the low deposition of dye indicated by the weak fluorescence behind sliding drops. Only for the strongly adsorbed cationic surfactant CTAB can partially remain on the solid surface (Figure 3, pathway 3).

In the case of cationic CTAB, the strong decrease in charge separation can be explained by the charge compensation. Even if only a small fraction of the positively charged CTAB remains on the solid surface (Figure 3, pathway 3), they compensate for the negative surface charges. To estimate how many CTAB molecules must be deposited to neutralize the primary anions, we assume a typical surface charge density of $-10\ \mu C/m^2$ deposited behind a water drop with ~1 nC drop charge [2, 18], which corresponds to $6\times10^{13}$ unit charges per $m^2$. Thus, one deposited CTAB cation every 130 nm is sufficient to neutralize the surface. Compared to the surface excess of CTAB at the CMC of the order of $10^{18}$ molecules/$m^2$ [36, 38], even a tiny fraction of the absorbed CTAB deposited on the solid surface can shield and even reverse the primary negative surface charge. We confirmed the low amount of deposited CTAB by the fact that the movement of the first drop is almost identical to that of subsequent drops within reasonable errors (Figure S5).

Of the three surfactants used, the nonionic $C_8E_3$ showed the least reduction in charge separation as compared to pure water. Surfactant tails shield the solid surface from primary ions, reducing the solid-liquid charge density. Simulations show that a 25% reduction in uncovered surface area reduces the maximum possible surface charge deposition by 10% (Figure S6).

Anionic SDS further reduces charge separation compared with nonionic $C_8E_3$. We attribute this to a reduction in the density of the primary ions. The negative charges of the adsorbed SDS prevent primary ions from accumulating at the solid-liquid interface. The increase in the surface potential was confirmed by the larger negative zeta potential (Figure 2a). According to the charge regulation model (SI 7) [18], surfactant absorption reduces the chemically bound surface charge at the solid-liquid interface. We speculate that a major effect reducing the density of primary ions to be deposited behind the drop is caused by the transfer of adsorbed surfactant to the liquid-air interface at the rear contact line. These surfactants can take primary ions with them.

Is there enough time for the surfactant to diffuse to the liquid-solid interface and cover a significant part of the interface? In a tilted plate experiment, a given area of the solid surface is only wetted for a short time. For example, a 6 mm long drop moving at 0.2 m/s will contact a given area for 30 ms. In contrast, all adsorption and zeta potential measurements were made over much longer time scales. To estimate how quickly an adsorption equilibrium is reached, we calculate the characteristic time for surfactant diffusion to an interface as $\tau = \Gamma_0^2/Dc^2$ [40-42]. Here, $\Gamma_0$ is the surface excess at equilibrium and $c$ is the concentration of the surfactant in the bulk. For example, for $C_8E_3$ at the CMC and $\Gamma_0 = 3.6 \times 10^{18}$ m$^{-2}$ [43] we get $\tau = 14$ ms. Thus, the surface excess is therefore close to equilibrium. However, at lower concentrations, $\tau$ is longer than typical contact times so the adsorbed amount does not reach equilibrium. Nevertheless, a significant amount of surfactant covers the solid-liquid interface at the rear of the drop.

Besides sliding drops, surfactants also screen slide electrification on bouncing drops. To measure the drop charge of bouncing drops with surfactants, we used a home-built setup reported before (Figure 4a) [44] (SI 6). Briefly, we recorded drop bouncing on soot-templated superhydrophobic surfaces from the side by a high-speed camera (Figure 4b, SI 2). An electric field is applied horizontally. Due to the net charge in the drop, the drop deflects from an initial position $(x_0, t_0)$ to the side $(x, t)$ after bouncing off the surface in the external electric field of field strength $E$. The drop charge ($q$) is quantified based on the electrostatic force ($F = qE = \rho Va$, $a$: acceleration, $V$: drop volume, $\rho$: drop density) for the deflection ($x - x_0 = \frac{1}{2}a(t-t_0)^2$). That is $q = \frac{2\rho V(x-x_0)}{E(t-t_0)^2}$. For a 5 μL water drop bounces on the soot-templated superhydrophobic surface, the drop charge is typically ~15 pC, consistent with our previous report [44]. The less charge of bouncing drops on superhydrophobic surface than sliding drops on hydrophobic surface is due to much smaller drop size and lower contact area. The deflection of bouncing drops decreased with increasing CTAB concentration (Fig. 4b), demonstrating a decrease in drop charge (Fig. 4c). The decrease of drop charge with surfactant concentration is in line with the trend of sliding drops. The reduction of drop charge for bouncing drop also occurs with 10% CMC SDS and 10% $C_8E_3$ (Fig. 4d). The reduction of bouncing drop charge is comparable between all three different surfactants. We attribute this to the influence of the applied electric filed on the absorption of surfactant during measurements and the different surface topography of superhydrophobic surfaces compared to flat hydrophobic surfaces. Therefore, the screen effect of surfactant on slide electrification is universal.

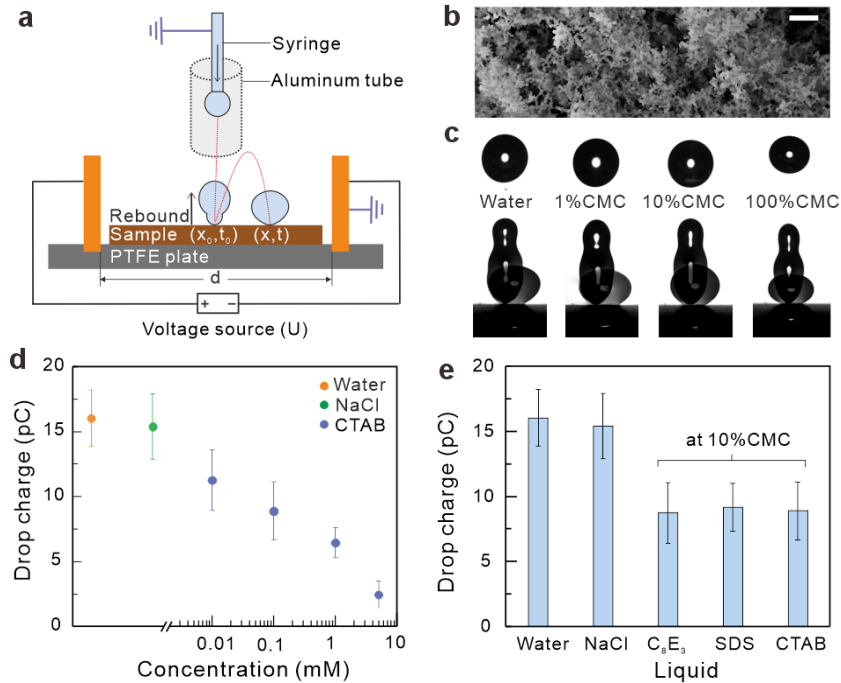

**Figure 4.** *Screen effect of surfactant for bouncing drops. (a) The setup for charge measurement by bouncing drop. $x_0$ and $t_0$ are the position and time when the bouncing drop first leaves the surface. $x$ and $t$ are the position and time when the drop contacts the surface again. $U$ is the applied voltage between the two parallel copper plate with distance $d$. Aluminum tube is used to prevent drop polarization. (b) Scanning force microscopy image of a candle soot templated superhydrophobic surface. (c) Image sequence of bouncing water drops with CTAB at different concentration in electric field of 100 KV/m. The syringe size is fixed. With increasing surfactant concentration, drop volume decrease from ~5 µL to ~3 µL. (d) Drop charge versus CTAB concentration after bouncing on superhydrophobic surfaces. (d) Drop charge of different liquid drops after bouncing on superhydrophobic surfaces.*

In conclusion, we found that surfactants screen slide electrification and its influence on drop motion. The screening effect occurs for both ionic and non-ionic surfactants resulting in the charge reduction. Cationic CATB reduces charge most due to charge compensation by the deposition of CTAB on the solid surface after dewetting. Anionic SDS and non-ionic $C_8E_3$ reduce charge less. We attribute the reduction by SDS and $C_8E_3$ to 1) their absorption shielding solid surface from primary ions at the solid-liquid interface and 2) their desorption further taking away primary ions from the solid-liquid interface to the liquid-air interface. The screen effect by surfactant might help to get rid of possible surface corrosion due to the high potential formed by slide electrification.

**Supporting Information**

The authors have cited additional references within Supporting Information [45-50].

**Acknowledgements**

We would like to thank Rüdiger Berger, Aaron D Ratschow, and Tobias Baier for the discussion. This work was supported by the European Research Council (ERC) under the European Union's Horizon 2020 research and innovation program (grant agreement no. 883631) (X.L., H.-J. B.). X.Z. and Z.N. would like to thank the China Scholarship Council (CSC) for the financial support.

**Conflict of interest**

The authors declare no conflict of interest

**Data Availability Statement**

The data that support the findings of this study are available in the supplementary material of this article.